\documentclass[10pt, a4paper]{article}

\usepackage{lrec-coling2024}

\usepackage{tempora}
\usepackage{listings}
\usepackage{subcaption}
\usepackage{multirow}
\usepackage{booktabs}
\usepackage{amsmath}
\usepackage{graphicx}

\usepackage{natbib}
\usepackage{multibib}
\makeatletter
\def\@mb@citenamelist{cite,citep,citet,citealp,citealt,citepalias,citetalias}
\makeatother
\newcites{languageresource}{~}

\usepackage{graphicx}
\usepackage{tabularx}
\usepackage{soul}
\usepackage{titlesec}
\titleformat{\section}{\normalfont\large\bfseries\center}{\thesection.}{1em}{}
\titleformat{\subsection}{\normalfont\SmallTitleFont\bfseries\raggedright}{\thesubsection.}{1em}{}
\titleformat{\subsubsection}{\normalfont\normalsize\bfseries\raggedright}{\thesubsubsection.}{1em}{}
\renewcommand\thesection{\arabic{section}}
\renewcommand\thesubsection{\thesection.\arabic{subsection}}
\renewcommand\thesubsubsection{\thesubsection.\arabic{subsubsection}}

\usepackage{xcolor}
\usepackage{hyperref}
\definecolor{darkblue}{rgb}{0, 0, 0.5}
\hypersetup{colorlinks=true, citecolor=darkblue, linkcolor=darkblue, urlcolor=darkblue}

\usepackage{xstring}

\usepackage{color}

\usepackage{multirow,booktabs,url}
\usepackage{acro}
\usepackage{colortbl}
\usepackage{float}
\restylefloat{table}
\newcommand{\scell}[2][c]{
\begin{tabular}[#1]{@{}c@{}}#2\end{tabular}}

\DeclareAcronym{asr}{
short = ASR,
long = automatic speech recognition,
}

\DeclareAcronym{cer}{
short = CER,
long = character error rate,
}

\DeclareAcronym{dtw}{
short = DTW,
long = dynamic time warping,
}

\DeclareAcronym{gpu}{
short = GPU,
long = graphics processing unit,
}

\DeclareAcronym{id}{
short = ID,
long = identifier,
}

\DeclareAcronym{mcd}{
short = MCD,
long = mel-cepstral distortion,
}

\DeclareAcronym{mos}{
short = MOS,
long = mean opinion score,
}

\DeclareAcronym{mfcc}{
short = MFCC,
long = mel-frequency cepstral coefficient,
}

\DeclareAcronym{rar}{
short = RAR,
long = relative attributes ranking,
}

\DeclareAcronym{sota}{
short = SOTA,
long = state-of-the-art,
}

\DeclareAcronym{ser}{
short = SER,
long = speech emotion recognition,
}

\DeclareAcronym{tts}{
short = TTS,
long = text-to-speech,
}

\title{KazEmoTTS:\\A Dataset for Kazakh Emotional Text-to-Speech Synthesis}

\name{Adal Abilbekov, Saida Mussakhojayeva, Rustem Yeshpanov, Huseyin Atakan Varol} 

\address{Institute of Smart Systems and Artificial Intelligence\\Nazarbayev University, Astana, Kazakhstan \\
\{adal.abilbekov, saida.mussakhojayeva, rustem.yeshpanov, ahvarol\}@nu.edu.kz\\}

\abstract{
This study focuses on the creation of the KazEmoTTS dataset, designed for emotional Kazakh text-to-speech (TTS) applications. KazEmoTTS is a collection of 54,760 audio-text pairs, with a total duration of 74.85 hours, featuring 34.23 hours delivered by a female narrator and 40.62 hours by two male narrators. The list of the emotions considered include ``neutral'', ``angry'', ``happy'', ``sad'', ``scared'', and ``surprised''. We also developed a TTS model trained on the KazEmoTTS dataset. Objective and subjective evaluations were employed to assess the quality of synthesized speech, yielding an MCD score within the range of 6.02 to 7.67, alongside a MOS that spanned from 3.51 to 3.57. To facilitate reproducibility and inspire further research, we have made our code, pre-trained model, and dataset accessible in our GitHub repository.
\\ \newline  
}

\begin{document}

\maketitleabstract

\section{Introduction}
The demanding challenges of generating high-quality synthesized speech for one or more speakers have been met by rapidly developed \ac{tts} systems~\citep{1712.05884,2006.04558,1705.08947}.
Yet, synthesized speech still faces significant difficulties in expressing paralinguistic features such as emotions. 

In the area of emotional \ac{tts}, where the voice synthesized by a \ac{tts} system is to convey emotions (e.g., anger, happiness, sadness, etc.), the availability of high-quality labeled datasets remains quite limited. 

As far as our knowledge extends, most publicly available emotional speech datasets primarily cover high-resource languages, such as Chinese, English, or French~\citep{adigwe2018emotional,costantini-etal-2014-emovo,10.1007/s10579-008-9076-6}. These datasets typically focus on either distinct emotional states (e.g., anger, happiness, etc.;~\citealt{costantini-etal-2014-emovo}) or emotion polarity, ranging from absolutely negative to absolutely positive~\citep{2106.09317}. They often include several narrators' speech samples and exhibit variations in terms of total audio duration~\cite{esd,2106.09317}.

In our study, we have undertaken the pioneering task of creating an emotional \ac{tts} dataset for Kazakh---a low-resource language. However, its utility extends beyond this specific domain and can be applied effectively in diverse areas, including \ac{ser} and emotional voice conversion tasks.
The dataset comprises a total of 74.85 hours of recorded high-quality speech data featuring six distinct emotional categories. The Kazakh emotional TTS (KazEmoTTS) dataset is comprised of contributions from three professional narrators, with 34.23 hours of the data provided by a female narrator and 40.62 hours by two male narrators.
Additionally, we introduce a \ac{tts} model, trained on KazEmoTTS, with the capability to produce Kazakh speech reflecting six emotional expressions.
KazEmoTTS and the model are openly accessible for both academic and commercial purposes, operating under the provisions of the Creative Commons Attribution 4.0 International License in our GitHub repository.\footnote{\url{https://github.com/IS2AI/KazEmoTTS}\label{ft:github}}

The structure of the paper is as follows: \hyperref[sec2]{Section~2} offers an overview of previous research in emotional \ac{tts}. \hyperref[sec3]{Section~3} describes the construction of the dataset.
\hyperref[sec4]{Section~4} covers the experimental design and evaluation metrics.
\hyperref[sec5]{Section~5} provides a  presentation of the experimental results and a brief summary of the main findings. 
\hyperref[sec6]{Section~6} concludes the paper.

\section{Related Work}\label{sec2}

Previous studies into the complex relationships and interactions between distinct emotional states suggest that individuals can potentially experience a wide array of diverse emotions~\citep{873e1018-42da-31ed-a115-a32187b6d0ca,article}.
~\citet{plutchik2013theories} distill a set of eight fundamental emotions, including anger, anticipation, disgust, fear, joy, sadness, surprise, and trust, with other emotional states believed to arise from various combinations of these. 

That said, Paul Ekman's well-known theory of six basic emotions~\citep{ekman1992argument} proposes the existence of anger, disgust, fear, happiness, sadness, and surprise, and is frequently invoked in emotional \ac{tts} research~\citep{Schröder2009,zhou2020converting,Zhou2022SpeechSW}.

Most datasets employed for emotional \ac{tts} include emotion labels, in contrast to prosody modeling approaches~\citep{2105.13086,2202.07200}, which do not rely on preset labels. Presently, emotional \ac{tts} research primarily revolves around two methods: (1) synthesizing speech with explicit, predefined emotional labels and (2) regulating the intensity of emotions in speech synthesis.

Employing hard-labeled emotions is generally considered the most straightforward approach.~\citet{1711.05447} utilized an attention-based decoder that captures an emotion label vector to generate the desired emotional style in the synthesized speech. 
In~\citet{2104.00436}, style embeddings were extracted from both a reference speech sample and a corresponding style tag.

With respect to models that allow for the control of emotional intensity, the prevailing method of determining emotional intensity is \ac{rar}~\citep{inproceedings}. RAR involves the creation of a ranking matrix that is derived through a max-margin optimization problem typically addressed using support vector machines. The solution is subsequently employed for model training. However, it is important to note that this process is manually constructed and therefore can potentially introduce biases into the training process~\citep{Guo2022EmoDiffIC}.

In~\citet{Um2019EmotionalSS}, the researchers introduced an algorithm designed to increase the gap between emotion embeddings. They also employed the interpolation of this embedding space as a means to control the intensity of emotions.
In~\citet{Im2022EMOQTTSEI}, quantization techniques were introduced to measure the distances between emotion embeddings, enabling the control of emotion intensities.

Similar methods have been applied to intensity control in emotion conversion~\citep{Choi2021SequencetoSequenceEV,Zhou2022EmotionIA}. However, even though an autoregressive model~\citep{Zhou2022SpeechSW} that relies on intensity values derived from \ac{rar} to weigh emotion embeddings is implemented, the problem of speech quality degradation persists.

EmoDiff~\citep{Guo2022EmoDiffIC}, built on the design of GradTTS~\cite{Popov2021GradTTSAD}, introduces a soft-label guidance approach inspired by the classifier guidance technique, employed in diffusion models~\citep{Dhariwal2021DiffusionMB,Liu2021MoreCF}.
The classifier guidance technique is a sampling method that leverages the gradient of a classifier to lead the sampling path when provided with a one-hot class label.
The adoption of an alternative approach can be observed in EmoMix~\citep{Tang2023EmoMixEM}, another GradTTS-based model. This approach combines a diffusion probabilistic model with a pre-trained \ac{ser} model. The emotion embeddings extracted by the \ac{ser} model act as an additional condition, enabling the reverse process within the diffusion model to generate primary emotions.

\section{Dataset Construction}\label{sec3}
\subsection{Text Collection}
Narration materials were drawn from multiple sources. Scientific, computer technology, historical, and international articles were retrieved from Kazakh Wikipedia. News content was collected from reputable Kazakh media outlets. In addition, selections from public domain books, fairy tales, and phrasebooks were included. All the collected texts were split sentence-wise.

\subsection{Recording Process}

We hired three professional narrators---one female and two males---for the project. The narrators were given the option to record in their personally arranged home studios or within the facilities of our institution. They were also given precise instructions to read the texts in quiet indoor settings while conveying a high degree of emotional expression in line with the specified emotions. 

Each sentence was paired with one of the six emotions selected for the study, ensuring an even distribution among all sentences.
Our selection of these emotions was informed by their prevalence in prior research~\citep{esd,adigwe2018emotional,10.1007/s10579-008-9076-6} and their discernibility by evaluators~\citep{costantini-etal-2014-emovo}. 
As a result, the list of emotions examined in our study included all those proposed in~\citet{ekman1992argument}, with the sole exception being ``disgust''. 
Additionally, we introduced a ``neutral'' option, representing the absence of a specific emotion.

\begin{figure}
\begin{center}
\subfloat[]{\label{collect}\includegraphics[scale=0.17]{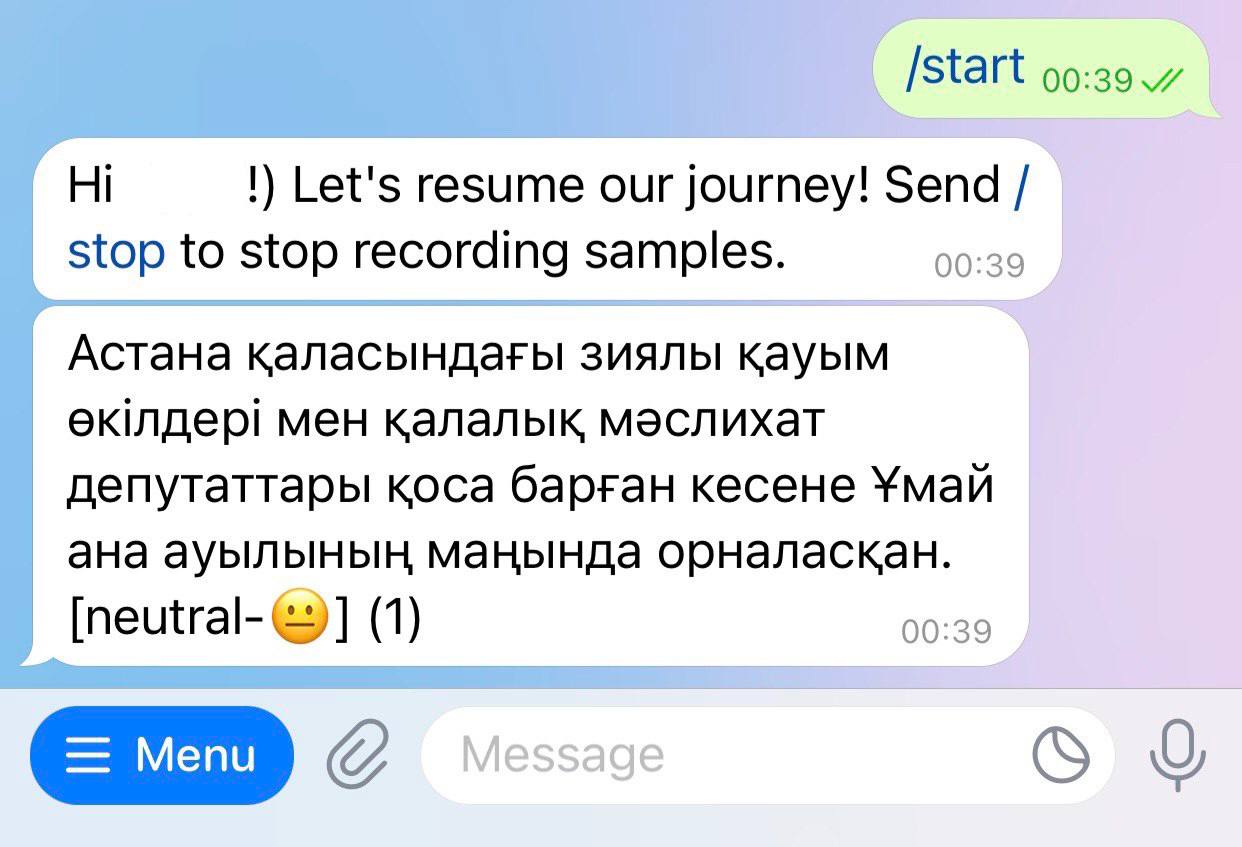}} \\
\subfloat[]{\label{eval}\includegraphics[scale=0.18]{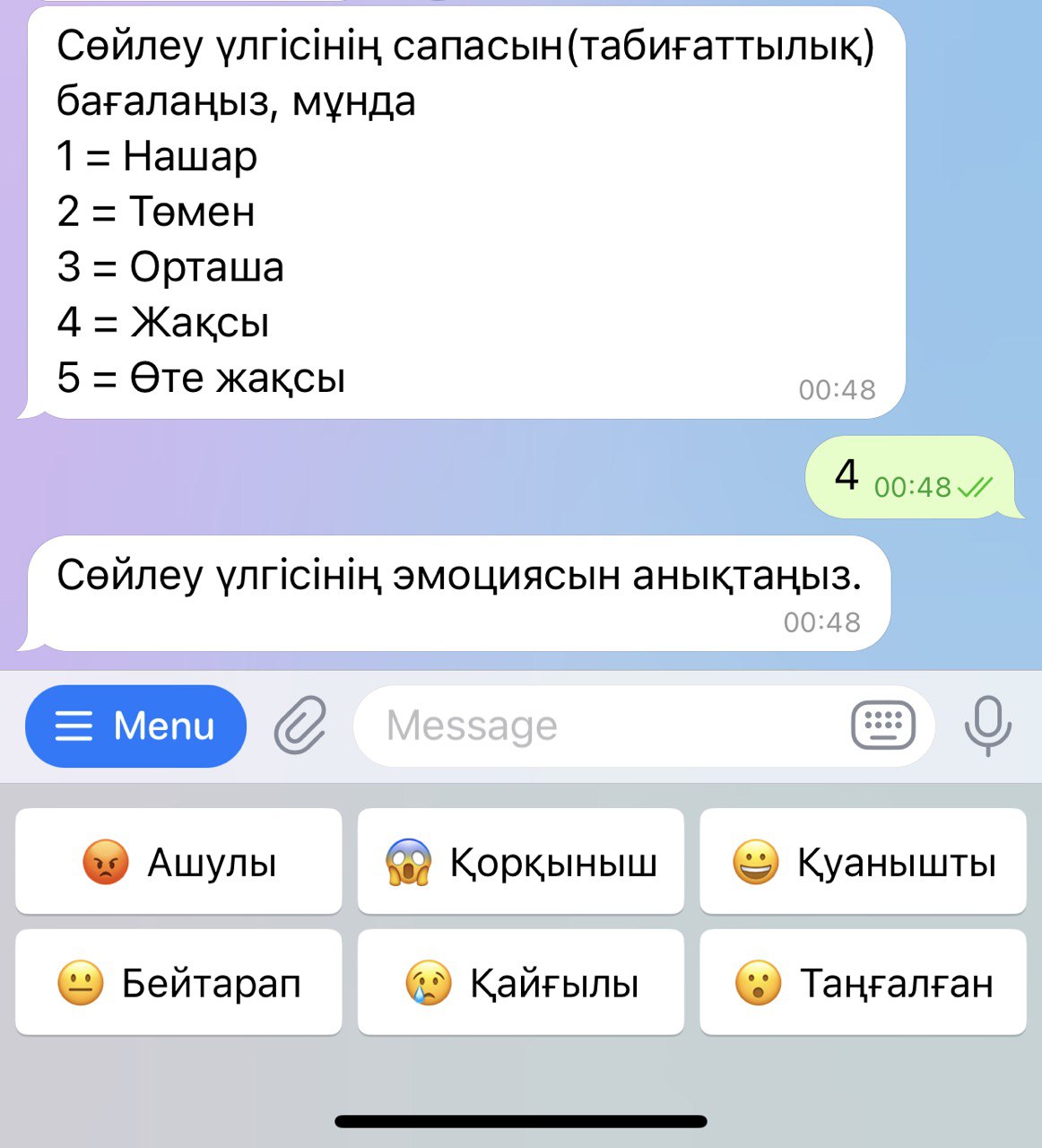}}
\caption{Telegram bot user interfaces: a) Narration functionality, and b) evaluation functionality.}
\label{fig_tg_bot}
\end{center}
\end{figure}

All recorded audio files were either sampled at a rate of 44.1 kHz and stored as 16-bit samples or at a rate of 48 kHz and stored as 24-bit samples. The whole data collection process was facilitated by a messenger application (Telegram) bot, as depicted in Figure~\ref{collect}.

\vspace{-0.1cm}

\begin{table*}[!ht]
\setlength\tabcolsep{0.17cm}
\begin{tabularx}\textwidth{cccccccccccccccc} 
\toprule
\multirow{2}{*}{\textbf{E}} & \multirow{2}{*}{\textbf{Count}} & \multicolumn{4}{c}{\textbf{F1}} &  & \multicolumn{4}{c}{\textbf{M1}} &  & \multicolumn{4}{c}{\textbf{M2}} \\
\cmidrule{3-6} \cmidrule{8-11} \cmidrule{13-16}
& & \textbf{Total} & \textbf{Mean} & \textbf{Min} & \textbf{Max} & & \textbf{Total} & \textbf{Mean} & \textbf{Min} & \textbf{Max} & & \textbf{Total} & \textbf{Mean} & \textbf{Min} & \textbf{Max} \\
\midrule
neu & 9,385 & 5.85 & 5.03 & 1.03 & 15.51 &  & 4.54 & 4.77 & 0.84 & 16.18 &  & 2.30 & 4.69 & 1.02 & 15.81 \\
ang & 9,059 & 5.44 & 4.78 & 1.11 & 14.09 &  & 4.27 & 4.75 & 0.93 & 17.03 &  & 2.31 & 4.81 & 1.02 & 15.67 \\
hap & 9,059 & 5.77 & 5.09 & 1.07 & 15.33 &  & 4.43 & 4.85 & 0.98 & 15.56 &  & 2.23 & 4.74 & 1.09 & 15.25 \\
sad & 8,980 & 5.60 & 5.04 & 1.11 & 15.21 &  & 4.62 & 5.13 & 0.72 & 18.00 &  & 2.65 & 5.52 & 1.16 & 18.16 \\
sca & 9,098 & 5.66 & 4.96 & 1.00 & 15.67 &  & 4.13 & 4.51 & 0.65 & 16.11 &  & 2.34 & 4.96 & 1.07 & 14.49 \\
sur & 9,179 & 5.91 & 5.09 & 1.09 & 14.56 &  & 4.52 & 4.92 & 0.81 & 17.67 &  & 2.28 & 4.87 & 1.04 & 15.81 \\
\bottomrule
\multicolumn{16}{l}{\textit{Note.} Each emotion (E) was abbreviated to its first three letters due to space constraints.}
\end{tabularx}
\vspace{-0.25cm}
\caption{\label{kaz_emo_segment} Recording count and duration statistics: Total (hours), Mean, Max, and Min (seconds)}
\end{table*}

\subsection{Audio-to-Text Alignment Verification}

We conducted a thorough examination of the recorded audio files using a customized version of the Whisper multilingual \ac{asr}~\citep{radford2022robust} system to assess the accuracy of the audio-to-text alignment. The \ac{asr} system generated transcriptions based on the audio files, which were subsequently compared with the original texts. Texts exhibiting a \ac{cer} were identified and subjected to a review process by a team of moderators. Recordings were excluded if they contained mispronunciations or significant background noise.

\subsection{Dataset Specification}

The audio recordings and their corresponding transcriptions are organized into separate folders for each narrator. 
All audio recordings were downsampled to a rate of 22.05 kHz and saved in the WAV format with a 16-bit per sample configuration.
They also underwent a preprocessing step involving the removal of silence and normalization, achieved by dividing the audio by its maximum absolute value.
The transcripts were saved as TXT files encoded in UTF-8 variable-length character encoding standard.
Both the audio and transcript files share identical filenames, differing only in their file extensions. Each file name comprises the narrator's \ac{id}, emotion, and utterance \ac{id}, structured as \textit{narratorID\_emotion\_utteranceID}.

The dataset contains 8,794 unique sentences and 86,496 unique words with an average sentence length of 10.83 words. Initially, a total of 84,714 audio files were recorded, but following quality checks, the dataset now contains 54,760 audio recordings. These recordings collectively represent an overall duration of 74.85 hours. 
The duration for the female narrator (F1) is 34.23 hours, with an average segment length of 5.0 seconds. The duration for the first male narrator (M1) is 26.51 hours, with an average audio segment length of 4.8 seconds. For the second male narrator (M2), the duration is 14.11 hours, with an average segment length of 4.9 seconds. 
More detailed statistics for the dataset are provided in Tables~\ref{kaz_emo_segment} and~\ref{tab:narrator_stats}.

\begin{table}[!h]
\centering
\begin{tabular}{ccc}
\toprule
\textbf{Narrator} & \textbf{\# recordings} & \textbf{Duration (h)}\\
\midrule
F1 & 24,656 & 34.23 \\
M1 & 19,802 & 26.51 \\
M2 & 10,302 & 14.11 \\

\midrule
\textbf{Total} & 54,760 & 74.85 \\
\bottomrule
\end{tabular}
\vspace{-0.25cm}
\caption{Narrator statistics}
\label{tab:narrator_stats}
\end{table}

\section{Experimental Setup}\label{sec4}
\subsection{KazEmoTTS Architecture}

We built our \ac{tts} model based on the design of GradTTS with hard label emotions~\citep{Popov2021GradTTSAD}, as was done in~\citet{Guo2022EmoDiffIC} and~\citet{Tang2023EmoMixEM}. The model was trained with the Adam optimizer and a learning rate of $10^{-4}$ for 3.7 million steps on one \ac{gpu} on an NVIDIA DGX A100 machine.
To improve the performance of diffusion models~\citep{Song2020ScoreBasedGM}, we implemented exponential moving averages for the weights of the model during training.
In addition, we removed the dependency on ground truth duration data and adjusted the sampling rate to 22,050 Hz.
During the inference phase, we set the guidance level parameter $\gamma$ to 100. The output of the model was an array of 80-dimensional log mel-filter bank features, representing acoustic features. To transform these acoustic data into time-domain waveform samples, we utilized the HiFiGAN vocoder~\citep{Kong2020HiFiGANGA}. Specifically, we trained it as a multi-speaker vocoder on the KazEmoTTS dataset without providing emotional labels for 1.72 million steps.
\subsection{Objective Evaluation}

We employed \ac{mcd}~\citep{Kubichek1993MelcepstralDM} as an objective assessment metric to evaluate the quality of the synthesized speech. This approach involves comparing the \ac{mfcc} vectors extracted from the generated speech and the ground truth speech, with a lower \ac{mcd} score suggesting that the generated speech is more similar to the ground truth.
To mitigate issues arising from the potentially extreme scaling of \ac{mcd} due to variations in the two input speech lengths, we adopted the \ac{dtw} algorithm, as described in~\citet{Battenberg2019LocationRelativeAM}. 

\subsection{Subjective Evaluation}
To evaluate the quality of the synthesized speech, we conducted a subjective evaluation survey via a messenger application (Telegram) bot. The user interface of the bot was developed in Kazakh, as shown in Figure~\ref{fig_tg_bot}b. 
To recruit volunteer participants, we distributed the link to the survey on popular social media platforms.

The survey involved a two-fold evaluation process. Participants were first tasked with evaluating the naturalness of a given speech sample, focusing on its degree of human-likeness. The evaluation was conducted using a five-point scale: 1.~\textit{bad},  2.~\textit{poor}, 3.~\textit{fair}, 4.~\textit{good}, and 5.~\textit{excellent}.
Subsequent to the naturalness evaluation, participants were prompted to identify one of the six distinct emotions with which the speech sample was narrated.

We compiled an evaluation set of 3,600 audio samples that were not included in the training set, from which a random subset of 36 (18 ground truth and 18 synthesized) speech samples was presented to each participant.
The samples were selected to ensure an equal representation of each narrator and emotion, amounting to six samples per narrator and one per emotion.
Participants were presented with one speech sample at a time. 
While participants were afforded the opportunity to listen to each sample multiple times, it was emphasized that their selection could not be altered once submitted.

\section{Results and Discussion}\label{sec5}
The evaluation results are provided in Tables~\ref{tab:mcd_and_mos_results}–~\ref{tab:emotion_perception}. As can be seen from Table~\ref{tab:mcd_and_mos_results}, on average, the synthesized speech delivered in a female voice demonstrated a greater likeness to the corresponding ground truth samples compared to the synthesized samples in both male voices.
An interesting observation is that synthesized speech samples featuring emotional states typically associated with lower-pitched voices (e.g., neutral, sad, scared) exhibited greater similarity to the corresponding ground truth samples. Conversely, speech samples generated to convey emotional states characterized by higher-pitched voices (e.g., angry, happy, surprised) demonstrated a comparatively lower degree of similarity to the ground truth samples.

\begin{table}[!h]
\fontsize{9}{10.8}\selectfont
\setlength\tabcolsep{0.115cm}
\begin{tabular}{c|ccccccc|cc}
\toprule
\multirow{2}{*}{\textbf{N}} & \multicolumn{7}{c}{\textbf{\ac{mcd}}} & \multicolumn{2}{|c}{\textbf{\acs{mos}}} \\
\cmidrule{2-10}
& neu & ang & hap & sad & sca & sur & \textbf{Avg} & GT & Syn \\
\midrule
F1 & 5.72 & 6.08 & 6.2 & 5.82 & 5.97 & 6.34 & \textbf{6.02} & 3.94 & 3.55 \\
M1 & 7.63 & 7.98 & 7.88 & 7.19 & 7.68 & 7.65 & \textbf{7.67} & 3.95 & 3.51 \\
M2 & 6.85 & 7.56 & 7.57 & 6.83 & 7.13 & 7.54 & \textbf{7.24} & 4.22 & 3.57 \\
\bottomrule
\multicolumn{10}{l}{\scell{\textit{Note.} N: narrators, Avg: average, GT: ground truth,\\Syn: synthesized}}
\end{tabular}
\caption{\ac{mcd} and \acs{mos} results}
\vspace{-0.25cm}
\label{tab:mcd_and_mos_results}
\end{table}

As for the evaluation survey, there were a total of 64 participants.
The \ac{mos} for assessing the naturalness of the synthesized speech varied only slightly, ranging from 3.51 to 3.57 (see Table~\ref{tab:mcd_and_mos_results}).
Narrator M2's samples attained the highest \acp{mos} for both the ground truth and synthesized samples, with M2's synthesized samples achieving a slightly higher score than that of F1 by a margin of 0.02.
The generated speech of Narrator M1 received the lowest \ac{mos}.
This underscores the lack of a correlation between \acs{mos} and the volume of data accessible for each narrator.
Despite the greater volume of data for Narrator F1 in comparison to the other two narrators, it did not translate into a much higher \ac{mos} for the female narrator. Similarly, the \ac{mos} was not higher for Narrator M1, despite having nearly twice as many data as Narrator M2.

A comparison of \acp{mos} highlights notably higher results in a separate study focused on Kazakh \ac{tts}~\citep{2201.05771}. For female speakers, ground truth speech evaluations achieved scores within the range of 4.18 to 4.73, while \acp{mos} for the generated speech covered a spectrum from 4.05 to 4.53. In the case of male speakers, \acp{mos} for ground truth ranged from 4.37 to 4.43, while scores for synthesized speech spanned from 3.95 to 4.2.

In English emotional \ac{tts} studies, results closely aligned with our findings were reported by \citet{Zhou2022SpeechSW}. They achieved a \ac{mos} of 3.45 when the emotion ``surprised'' was presented with 0\% intensity of other emotions. Notably, superior performance was observed with EmoDiff~\citep{Guo2022EmoDiffIC}, scoring 4.01, and EmoMix~\citep{Tang2023EmoMixEM}, which attained a \ac{mos} of 3.92.

As illustrated in Table~\ref{tab:accuracy}, sentences delivered in a neutral manner were more accurately recognized as such, achieving an accuracy rate of 65\%. In contrast, sentences expressed with anger proved to be the most challenging to identify, with a recognition accuracy of only 22\%.

\begin{table}[!ht]
\fontsize{8}{9.6}\selectfont
\setlength\tabcolsep{0.11cm}
\begin{tabularx}\columnwidth{c|cc|cc|cc|ccc}
\toprule
\multirow{3}{*}{\textbf{E}} & \multicolumn{2}{c|}{\textbf{F1}} & \multicolumn{2}{c|}{\textbf{M1}} & \multicolumn{2}{c|}{\textbf{M2}} & \multicolumn{3}{c}{\textbf{F1 \& M1 \& M2}} \\
\cmidrule{2-10}
& \textbf{GT} & \textbf{Syn} & \textbf{GT} & \textbf{Syn} & \textbf{GT} & \textbf{Syn} & \textbf{GT} & \multicolumn{1}{c}{\textbf{Syn}} & \textbf{Overall} \\
\midrule
neu & 0.75 & 0.60 & 0.64 & 0.67 & 0.53 & 0.67 & 0.64 & 0.65 & 0.65\\
ang & 0.28 & 0.07 & 0.26 & 0.06 & 0.58 & 0.07 & 0.37 & 0.07 & 0.22\\
hap & 0.59 & 0.35 & 0.50 & 0.32 & 0.64 & 0.41 & 0.58 & 0.36 & 0.47\\
sad & 0.22 & 0.32 & 0.40 & 0.29 & 0.37 & 0.26 & 0.33 & 0.29 & 0.31\\
sca & 0.25 & 0.35 & 0.42 & 0.29 & 0.28 & 0.35 & 0.32 & 0.33 & 0.33\\
sur & 0.32 & 0.19 & 0.28 & 0.23 & 0.33 & 0.20 & 0.31 & 0.21 & 0.26\\
\midrule
\textbf{Total} & 0.38 & 0.31 & 0.42 & 0.32 & 0.46 & 0.33 & 0.43 & 0.32 & 0.37\\
\bottomrule
\end{tabularx}
\caption{Results of emotion prediction accuracy}
\label{tab:accuracy}
\end{table}

\vspace{-0.2cm}

Table~\ref{tab:emotion_perception} displays the percentages of participant responses regarding their choice of emotion and reveals that ``neutral'' was frequently selected by participants when identifying the emotion of a speech sample. Interestingly, ``happy'' was the most easily identifiable emotion, chosen by nearly half of all participants, irrespective of the narrator. 
It is also worth noting that participants faced challenges when distinguishing ``angry'' speech samples, often mistaking them for ``sad'' or ``scared'' expressions. Frequently, samples labelled as ``scared'' were also erroneously identified as ``sad''.

\begin{table}[!ht]
\fontsize{10}{12}\selectfont
\setlength\tabcolsep{0.1cm}
\begin{tabularx}\columnwidth{ccc|c|c|c|c|c|c}
\toprule
& & \multicolumn{6}{c|}{\textbf{Participant responses (\%)}} & \multicolumn{1}{c}{\multirow{2}{*}{\textbf{N}}} \\
& & \multicolumn{1}{c}{\textbf{neu}} & \multicolumn{1}{c}{\textbf{ang}} & \multicolumn{1}{c}{\textbf{hap}} & \multicolumn{1}{c}{\textbf{sad}} & \multicolumn{1}{c}{\textbf{sca}} & \multicolumn{1}{c|}{\textbf{sur}} & \multicolumn{1}{c}{}\\
\multirow{18}{*}{\rotatebox[origin=c]{90}{\textbf{Actual emotions}}} & \multirow{3}{*}
{\rotatebox[origin=c]{90}{\textbf{neu}}} & 66.92 & 4.62 & 4.62 & 13.08 & 2.31 & 8.46 & F1 \\
& & 65.32 & 4.05 & 6.36 & 12.14 & 5.78 & 6.36 & M1 \\
& & 59.60 & 15.23 & 3.31 & 11.92 & 4.64 & 5.30 & M2 \\
\cmidrule{3-9}
& \multirow{3}{*}{\rotatebox[origin=c]{90}{\textbf{ang}}} & 43.31 & 16.56 & 3.18 & 14.01 & 14.01 & 8.92 & F1 \\
& & 39.04 & 15.75 & 2.74 & 18.49 & 13.70 & 10.27 & M1 \\
& & 31.33 & 33.33 & 2.67 & 10.00 & 12.67 & 10.00 & M2 \\
\cmidrule{3-9}
& \multirow{3}{*}{\rotatebox[origin=c]{90}
{\textbf{hap}}} & 37.75 & 0.00 & 45.70 & 3.97 & 1.99 & 10.60 & F1 \\
& & 39.73 & 8.90 & 43.84 & 2.74 & 10.27 & 9.15 & M1 \\
& & 28.28 & 2.07 & 51.72 & 1.38 & 2.76 & 13.79 & M2 \\
\cmidrule{3-9}
& \multirow{3}{*}{\rotatebox[origin=c]{90}
{\textbf{sad}}} & 35.63 & 3.13 & 3.75 & 25.63 & 19.38 & 12.50 & F1 \\
& & 46.43 & 2.38 & 0.60 & 34.52 & 7.74 & 8.33 & M1 \\
& & 50.00 & 0.74 & 2.94 & 31.62 & 8.82 & 5.88 & M2 \\
\cmidrule{3-9}
& \multirow{3}{*}{\rotatebox[origin=c]{90}
{\textbf{sca}}} & 21.77 & 17.01 & 6.12 & 20.41 & 29.93 & 17.01 & F1 \\
& & 23.98 & 2.92 & 1.17 & 26.90 & 35.09 & 9.94 & M1 \\
& & 20.95 & 16.89 & 0.68 & 12.84 & 31.76 & 16.89 & M2 \\
\cmidrule{3-9}
& \multirow{3}{*}{\rotatebox[origin=c]{90}
{\textbf{sur}}} & 37.04 & 2.47 & 12.96 & 4.94 & 14.81 & 27.78 & F1 \\
& & 30.41 & 5.85 & 4.09 & 18.71 & 15.79 & 25.15 & M1 \\
& & 37.01 & 3.15 & 11.02 & 12.60 & 9.45 & 26.77 & M2 \\
\bottomrule
\multicolumn{9}{l}{\textit{Note.} N: narrators.}
\end{tabularx}
\caption{Emotion perception results}
\vspace{-0.25cm}
\label{tab:emotion_perception}
\end{table}

In light of the \ac{mos} results acquired, it is apparent that despite providing participants with explicit instructions to focus on the emotion conveyed by the delivery of a sentence, rather than its inherent message or content, we acknowledge the cognitive challenge of perceiving a sentence with inherently somber content, such as one related to a grave illness or loss of life, being articulated with a seemingly cheerful emotion. This presents a cognitive challenge that may not have been fully resolved in our current study. In our future emotional \ac{tts} studies, we aim to address this challenge more effectively.

\section{Conclusion}\label{sec6}
This study aimed to construct the KazEmoTTS dataset for Kazakh emotional \ac{tts} applications. The dataset comprises a substantial 54,760 audio-text pairs, covering a total duration of 74.85 hours. This includes 34.23 hours delivered by a female narrator and 40.62 hours by two male narrators. The emotional spectrum within the dataset covers ``neutral'', ``angry'', ``happy'', ``sad'', ``scared'', and ``surprised'' states. In addition, a \ac{tts} model was developed through training on the KazEmoTTS dataset. Both objective and subjective evaluations were performed to gauge the synthesized speech quality, resulting in an objective \ac{mcd} metric ranging from 6.02 to 7.67 and an \ac{mos} ranging from 3.51 to 3.57. Our findings are particularly promising, considering that this study represents the first attempt at emotional \ac{tts} for Kazakh. To facilitate replicability and further exploration, we have made our code, pre-trained model, and dataset available in our GitHub repository.\textsuperscript{\ref{ft:github}}

\section{Acknowledgements}
We would like to express our gratitude to the narrators for their contributions and to the anonymous raters for their valuable evaluations.

\section{Bibliographical References}\label{references}
\bibliographystyle{references_style}
\bibliography{references}

\end{document}